\documentclass[prl]{revtex4}
\pdfoutput=1
\usepackage[utf8]{inputenc}
\usepackage{graphicx}
\makeatletter
\renewcommand{\@biblabel}[1]{\quad #1.}
\makeatother
\begin{document}
% Definition of title page:
%\linenumbers

\title{Unifying impacts in granular matter from quicksand to cornstarch
}
%From liquefaction to transient solidification during impact on jammed suspensions
%Unifying impacts in granular suspensions from quicksand to cornstarch

%From liquefaction to transient solidification during impact on fluid-saturated granular media.
%Transition from liquefaction to transient solidification during impact on fluid-saturated granular media. 
%Transition from liquefaction to transient solidification during impact on wet granular media.
%Transition from liquefaction to transient solidification during impact on jammed suspensions. fluid-saturated granular media. 
%From liquefaction to transient solidification during impact on jammed suspensions

\author{J. John Soundar Jerome$^{1,2\dagger}$, Nicolas Vandenberghe$^2$, Yo\"{e}l Forterre$^1$}  % insert author(s) here
\affiliation{1) Aix--Marseille Universit\'{e}, CNRS, IUSTI UMR $7343$, $13453$ Marseille Cedex $13$, France. 2) Aix--Marseille Universit\'{e}, CNRS, Centrale Marseille, IRPHE UMR $7342$, $13384$ Marseille Cedex $13$, France. $\dagger$ Present address: Universit\'{e} Claude--Bernard Lyon 1, CNRS, LMFA  UMR 5509, F-69622 Villeurbanne Cedex 08, France}

\begin{abstract}
A sharp transition between liquefaction and transient solidification is observed during impact on a granular suspension depending on the initial packing fraction. We demonstrate, via high-speed pressure measurements and a two-phase modeling, that this transition is controlled by a coupling between the granular pile dilatancy and the interstitial fluid pressure generated by the impact. Our results provide a generic mechanism for explaining the wide variety of impact responses in particulate media, from dry quicksand in powders to impact-hardening in shear-thickening suspensions like cornstarch.

\end{abstract}
\vskip 1.5cm
\pacs{???}
\maketitle

% optional
%\small{}{ {\fontsize{8}{10pt} }
Impacts on particulate media like granular materials and suspensions  present an astonishingly rich phenomenology  \cite{RuizSuarez_2013,Omidvar_2014}. Along with its astrophysical \cite{Melosh_book} and ballistics applications \cite{Wagner_2009}, impact dynamics is an object of active research to understand the high-speed response of granular matter \cite{Clark_PRL2015}. In dry granular media, impact by a solid object results in the formation of a corona of granular ejecta and a solid--like plastic deformation leading to a permanent crater \cite{Walsh_PRL2003,Uehara_2003,Yamamoto_2006,Deboeuf_2009,Marston_JFM2012,Zhao_PNAS2015}. For fine powders in air, granular jets and cavity collapse occur during impact \cite{Thoroddsen_2001, Lohse_2004}.  Subsequent studies showed that the ambient pressure of the interstitial fluid (air) is an important element for the observed fluid--like behavior  \cite{Royer_Nature2005,Caballero_PRL2007,Royer_2008}, while for denser packing the impact penetration is much reduced  \cite{Royer_EPL2011}. However, the question of the physical mechanisms and control parameters that give rise to such a wide variety of phenomena is still largely open.  Recently, studies on shear--thickening suspensions (cornstarch) showed completely different behaviors. Above a critical velocity, an impacting object immediately stops \cite{Waitukaitis_2012}, or in some cases generates cracks  \cite{Roche_PRL2013}, as if hitting a solid. This phenomenon has been related to the propagation of dynamic jamming  fronts in the bulk \cite{Waitukaitis_2012} but the mechanism remains unclear and overlooks the role of fluid/grains couplings, which are known to strongly affect the transient behavior of saturated granular materials \cite{Iverson_RevGeoPhy1997,Pailha_JFM2009,Andreotti_Forterre_Pouliquen_2013}. Whether impact-activated solidification relies on such couplings or on the complex rheology of the suspension is a pivotal question for clarifying the physics of shear-thickening fluids~--~a still highly debated topic \cite{Fernandez_2013,Seto_2013,Brown_2014,Wyart_2014,Fall_2015}.

The objective of this Letter is to address these questions and elucidate the role of the interstitial fluid and the initial volume fraction on the diverse impact phenomenology observed in granular materials and dense suspensions during the last decade. To avoid difficulties associated with colloidal interactions between particles (like in shear-thickening suspensions) or fluid compressibility (like in powders in air), we study here the impact of a freely-falling rigid sphere on a simple granular suspension \cite{Boyer_PRL2011} made up of macroscopic, heavy particles (glass beads in the range 0.1--1 mm) immersed in an incompressible liquid (water, viscous oil). The initial packing fraction of the suspension $\phi_0$ (the ratio of the volume of the glass beads to the total volume) is controlled by first fluidizing the mixture and then compacting the sediment, before removing the excess liquid  (see Fig. 1(a) and Supplemental Materials for the detailed set-up). For a granular pile of frictional spherical particles,  $\phi_0$ typically takes values between 0.55 and 0.62 \cite{Andreotti_Forterre_Pouliquen_2013}. The suspension is kept fully saturated so that surface tension effects can be neglected. 

 \begin{figure*}[t]
\begin{centering}
\includegraphics[width=17cm]{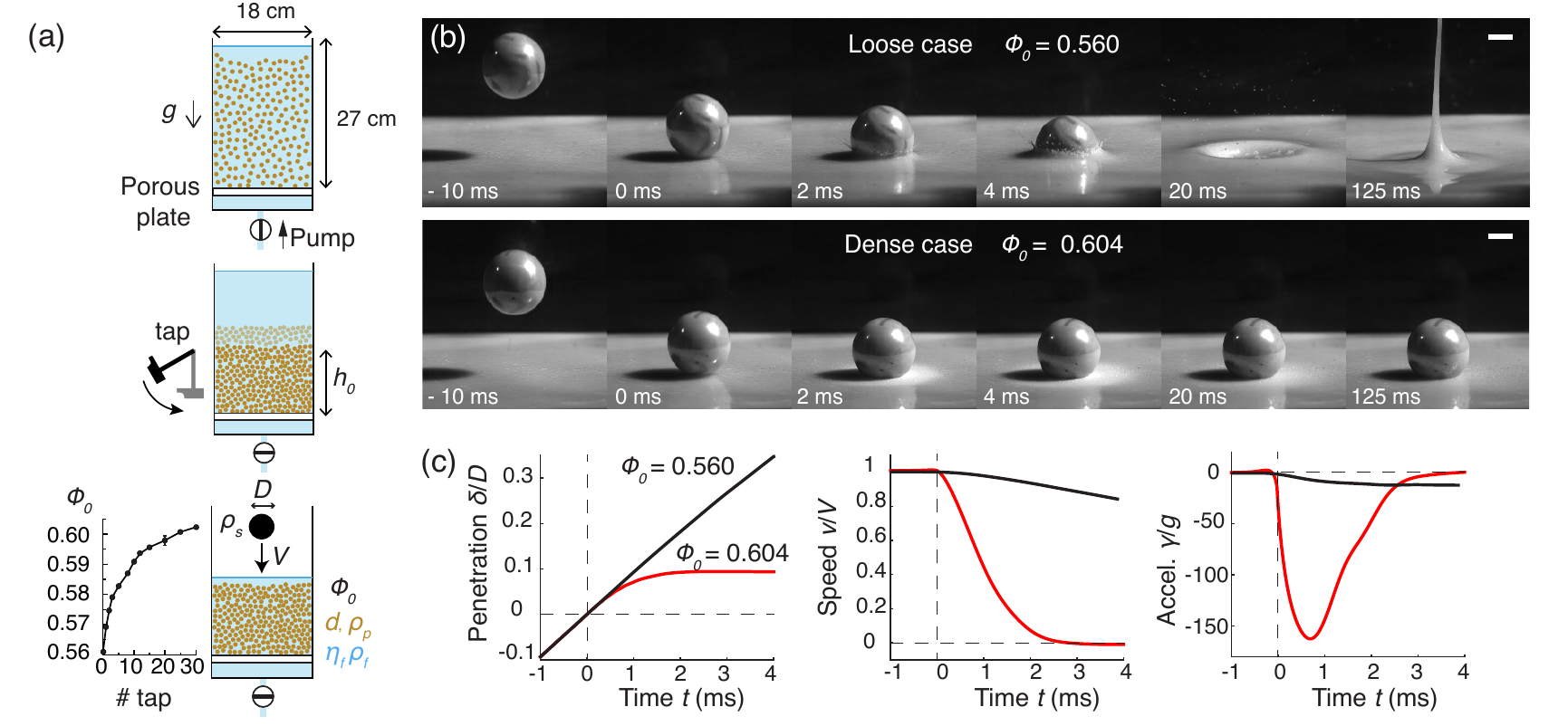}
\caption{Liquid-solid transition after impact on a suspension of heavy particles  (glass beads of diameter $d=170$ $\mu$m in water). (a) Protocol used to prepare the non-buoyant suspension at a given initial packing fraction $\phi_0$ ($h_0=9-10$ cm depending on the number of taps). (b) Image sequence of a solid sphere (glass marble of diameter $D=25.2$ mm, density $\rho_s=2.5$ g cm$^{-3}$, speed $V=2.35$ m s$^{-1}$) impacting the suspension in the loose (top) or dense (bottom) case. (c) Corresponding penetration, speed and acceleration of the impacting sphere as a function of time in milliseconds. Time $t=0$ gives the instant when the sphere hits the granular bed surface. Scale bars, 1 cm.}
\label{fig:SetUp}
\end{centering}
\end{figure*}

Remarkably, two very distinct impact regimes are observed depending on the initial packing fraction  [Fig. 1(b)].  For initially loose packing ($\phi_{0} = 0.560$), the ball readily sinks in the suspension, giving rise to a collapsing cavity and a central jet (see movie 1 in Supplemental Material). This is the typical behavior of a sphere impacting a liquid pool wherein the dynamics is dominated by fluid inertia \cite{Wor1908}. Such fluid-like behavior is also strikingly similar to that observed when a rigid sphere hits a dry loose powder under atmospheric pressure \cite{Thoroddsen_2001, Lohse_2004}.  By contrast, for dense packing ($\phi_{0} = 0.604$), the ball stops abruptly within a few milliseconds as it hits the surface (see movie 2 in Supplemental Material), with a huge deceleration of about $150$ $g$  (where $g =9.81$ m s$^{-2}$)   [Fig. 1(c)]. This solid-like  behavior is strongly reminiscent of the impact--activated solidification observed in shear-thickening suspension like cornstarch \cite{Waitukaitis_2012}. It also clearly depends on the grain size and viscosity of the interstitial fluid between the particles.  Using coarser particles in the same fluid tends to suppress the extreme deceleration, which can be restored by increasing the fluid viscosity (see movie 3 in Supplemental Material). The impact dynamics is also very different when the same grains are put in air instead of water. In this case, as reported in previous studies \cite{Walsh_PRL2003,Yamamoto_2006,Marston_JFM2012}, grains ejecta followed by the formation of a permanent crater is observed (see movie 4 in Supplemental Material). Therefore both the initial packing fraction and the interstitial fluid play a key role on the suspension behavior during impact.

To explain how such a drastic change in behavior can occur with only a slight  change of packing fraction ($\sim 5 \%$), we rely on a pore-pressure feedback mechanism: a coupling between the deformation of the granular medium and the pressure of the interstitial fluid between the grains \cite{Iverson_RevGeoPhy1997,Pailha_JFM2009}. As first described by O. Reynolds \cite{Reynolds_1885}, when a dense granular packing starts to flow, it must dilate. Since the medium is saturated with an incompressible liquid, the fluid is sucked in, as evidenced by the bright zone developing beneath the impacting sphere in the dense case [Fig. 1(b)]. Therefore, when the ball hits the pile, the interstitial fluid pressure (pore--pressure) drops instantaneously,  which in turn  presses the grains against each other thereby  enhancing the friction. Thus, the medium is transiently solidified.   A loosely packed granular bed, on the other hand, tends to compact when it deforms. Therefore, a rise in pore--pressure is produced during impact that can balance the weight of the grains. This suppresses the contact network, resulting in local or global fluidization. 
 
 We proceed a direct verification of this mechanism by developing a high-frequency measurement of the interstitial fluid pressure inside the suspension, just under the impact [Fig. 2(a)] (see Fig . S1 and Supplemental Material for the calibration procedure). In the loose case [Fig. 2(b), upper panel], the pore--pressure shows a sudden positive peak after impact. The measured peak pressure ($\sim 1$ kPa) is larger than the effective weight of the suspension above the sensor, $\phi_0 \Delta \rho g z\approx 400$ Pa ($\phi_0= 0.56$, $z=5$ cm, and $\Delta \rho = 1500$ kg m$^{-3}$), indicating a fluidization of the medium. At longer times sedimentation proceeds and the pressure slowly relaxes. By contrast, in the dense case [Fig. 2(b) lower panel], the peak pressure is negative and its magnitude ($10-100$ kPa) is much higher than the confining pressure due to gravity, indicating that particles are strongly pressed against each other, effectively leading to a solidification of the medium. Interestingly, the perturbation triggered by the impact remains localized near the impact point as shown by the pressure profile along the $z$-direction within the medium (inset of lower panel of Fig. 2b).  The decay length scale is $\sim$ 1 cm and much smaller than the container size, ensuring that the results reported are independent of wall effects. The transition between a positive and negative peak pressure occurs for a critical packing fraction $\phi_{c} = 0.585 \pm 0.0053$, which is independent of the projectile diameter $D$ and impact speed $V$ [Fig. 2(c)]. This value is consistent with previous rheological measurements of the jamming packing fraction using frictional spherical particles  \cite{Boyer_PRL2011}.

\begin{figure}
\begin{center}
\includegraphics{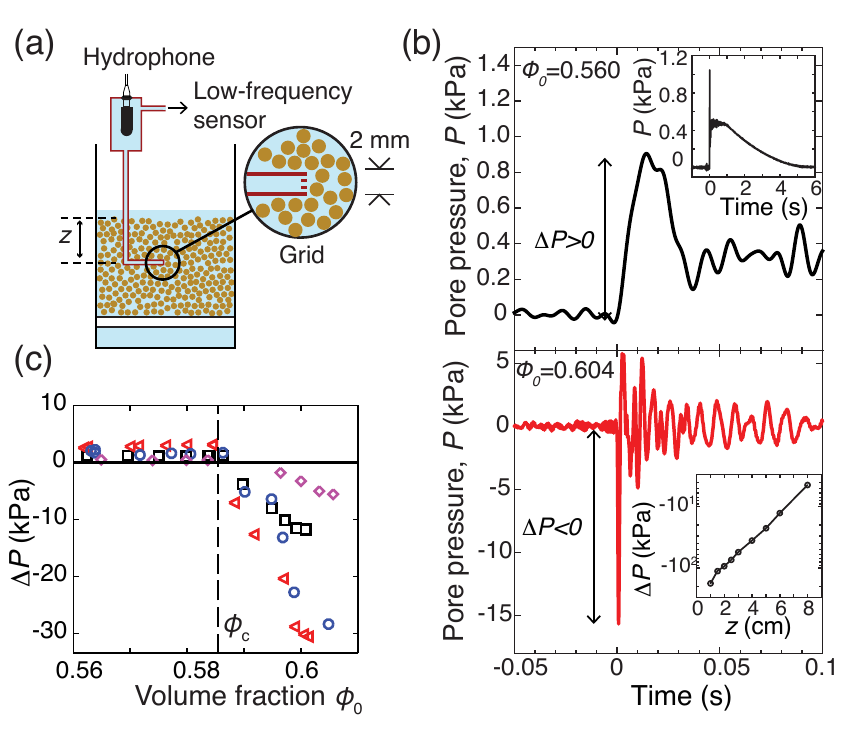}
\caption{Pore--pressure controls liquid-solid transition. (a) Experimental set-up to measure pore--pressure at different heights $z$ in the suspension. (b), Pore--pressure as function of time in the loose case (upper panel) and dense case (lower panel) (same experimental conditions as in Fig. 1, $z=6$ cm). In the dense case, the negative peak pressure intensity decreases with the distance from the impact (inset). (c), Peak pore--pressure as function of the initial packing fraction $\phi_0$ for various impact speeds $V$, sphere diameters $D$ and sensor positions $z$ (diamonds: $V=1.0$ m s$^{-1}$, $D = 16.5$ mm, $z= 2$ cm; 
circles: $V=3.4$ m s$^{-1}$, $D = 16.5$ mm, $z = 2$ cm;
square: $V=2.4$ m s$^{-1}$, $D = 25.2$ mm, $z = 5$ cm;  
triangles: $V=4.8$ m s$^{-1}$, $D = 25.2$ mm, $z = 5$ cm).    }
\label{fig:PorePressure}
\end{center}
\end{figure}

%*******version longue****** 

In order to model the impact dynamics, the coupling between the grain matrix deformation and the interstitial fluid should necessarily be taken into account. For an assembly of rigid particles, the simplest dilatancy law relating the evolution of the volume fraction $\phi$ and the packing deformation at the onset of plastic flow  is given by    \cite{Roux_Radjai_1998,Andreotti_Forterre_Pouliquen_2013}
\begin{equation}
\label{eq:Reynolds}
	\frac{1}{\phi}\frac{\partial\phi}{\partial t} = - \dot{\gamma} \tan\Psi=-\alpha \dot{\gamma}\left({\phi-\phi_c}\right),
\end{equation}
where  $\dot{\gamma} > 0$ is the absolute shear rate of the granular medium and $\Psi$ is the Reynolds dilatancy angle  \cite{Reynolds_1885}, assumed proportional to $\phi-\phi_c$, and  $\alpha$ a constant of order unity.  This deformation of the granular matrix induces in turn an interstitial fluid flow, which, for the  low Reynolds numbers considered here, is described by the Darcy law \cite{Darcy1856,Jackson_Book2000}
\begin{equation}
\label{eq:Darcy}
(1-\phi)(\textbf{V}_f-\textbf{V}_{p}) = - \frac{\kappa}{\eta_{f}} \mathbf{\nabla} P_{f},
 \end{equation}
where the vector fields $\textbf{V}_{p}$ and $\textbf{V}_f$ denote the particle and liquid velocity, respectively, $P_{f}$ is the liquid pore--pressure, $\eta_f$ the fluid viscosity and  $\kappa\propto d^2$ the permeability of the granular pile.  
Assuming that the liquid and the particles are incompressible: ${\partial \phi}/{\partial t} +\nabla \cdot ( \phi \textbf{V}_{p} ) = 0$ and ${\partial\left(1- \phi\right)}/{\partial t} +\nabla \cdot ( \left(1-\phi\right)\textbf{V}_f ) = 0$,  and taking the divergence of the Darcy law (\ref{eq:Darcy}) gives $(1/{\phi}) {\partial \phi}/{\partial t} = -({\kappa}/{\eta_{f}}) \nabla^2 P_{f}$, in which spatial variations of $\phi$ have been neglected.
Using the Reynolds dilatancy equation (\ref{eq:Reynolds}), we finally obtain a Poisson-like equation for the pore--pressure 
\begin{equation}
\label{eq:DarcyReynolds}
	\nabla^2 P_{f} = \frac{\eta_{f}}{\kappa} \alpha  \dot{\gamma} \left({\phi-\phi_c}\right). 
\end{equation}
in which the sign of the source term  is imposed only by $\phi-\phi_c$.  Therefore, the pore--pressure generated by the impact scales as
 \begin{equation}
\label{eq:DarcyReynoldsScaling}
P_{f} \sim - \frac{\eta_{f}}{\kappa} \alpha  \Delta \phi V_p L,
\end{equation}
where $\Delta \phi=\phi_0-\phi_c$, $V_p$ is the velocity scale for the particle velocity field, $L$ the typical extent up to which deformation is experienced by the granular bed, and $\alpha$ a constant of order unity. This relation predicts that the pore pressure is positive (fluid-like response) or negative (solid-like response) depending only on the sign of $\Delta \phi$, while its magnitude is controlled by both the grain diameter and fluid viscosity, in agreement with observations (Fig. 2c and movie 3 in the Supplementary Material).

The Darcy-Reynolds model (\ref{eq:DarcyReynoldsScaling}) can also be used to quantitatively infer the penetration dynamics in the dense case   ($\Delta \phi>0$). Assuming a frictional rheology for the granular suspension \cite{Andreotti_Forterre_Pouliquen_2013} and neglecting the confining pressure due to gravity in front of the pore--pressure, the contact stress on the impactor is $-A P_f$, where $A$ is an effective friction coefficient   \cite{Schofield_1968,Seguin_2013} and  $P_f$ is the pore--pressure (\ref{eq:DarcyReynoldsScaling}), in which $V_p=\dot{\delta}$, where $\delta$ is the penetration depth, and $L=a$ the typical radius of the contact area of the projectile (consistent with our measurements of the pressure profile, see inset of lower panel of Fig. 2b). Using Newton's second law for the impactor $\rho_s ({\pi D^3}/{6}) \ddot{\delta}=\pi a^2 A P_f$, where  $\rho_s$ is the density of the projectile,  and assuming small penetration ($a^2 \approx D\delta$), the penetration $\delta$ then evolves according to the non-dimensional equation (after integration with initial conditions $\delta (0) = 0$ and $\dot{\delta} (0) = V$): $d\tilde{\delta} / d \tilde{t} = - (2/5) \tilde{\delta}^{5/2}  + 1$, where $\tilde{\delta}=\delta/(Vt_m)$ is the dimensionless penetration and $\tilde{t}=t/t_m$ a dimensionless time given by
\begin{equation} \label{eq:timelength}
  t_m = \frac{D}{V} \left( \lambda \Delta\phi \right)^{-2/5} \mathrm{where} \ \lambda =  6  A \alpha \frac{\eta_{f} D} {\rho_s \kappa V }.
\end{equation}

\begin{figure}
\begin{center}
\includegraphics{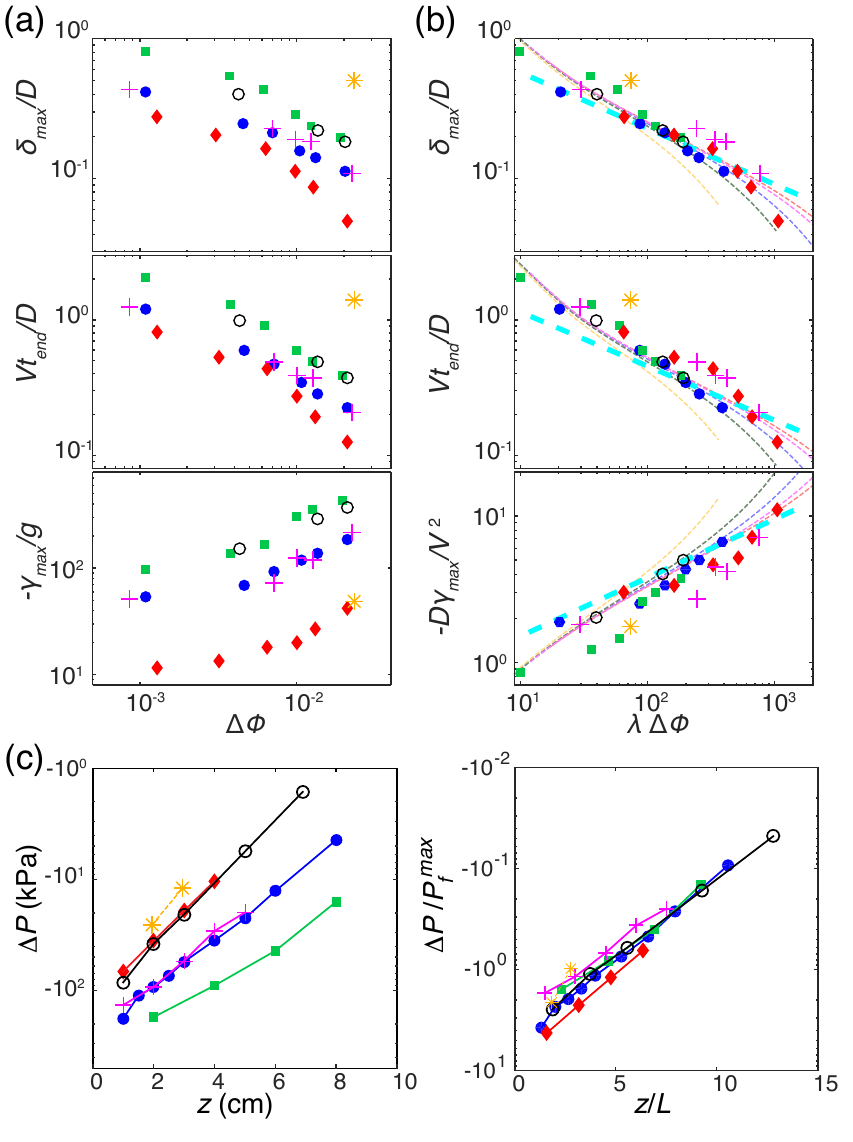}
\caption {Scaling laws for the penetration dynamics and pore--pressure in the dense case.  (a) The maximal indentation $\delta_{max}$, the typical stopping time $t_{end}$ (the time at which $\dot{\delta}/V = 0.05$) and the maximal acceleration $\gamma_{max}$ plotted against $\Delta \phi$ for different impact speeds (diamonds: $V = 1.0$ m s$^{-1}$, disks: $V = 2.5$ m s$^{-1}$, squares: $V = 5.3$ m s$^{-1}$, with $D = 25.2$ mm), impactor size (circles: $V = 3.5$ m s$^{-1}$, $D = 16.5$ mm) and suspension properties (diamonds, disks, squares, circles: particles $d = 170$ $\mu$m in water; crosses: particles $d = 500$ $\mu$m in a viscous fluid $\eta_f = 10.5\times 10^{-3}$ Pa$\,$s, stars:  particles $d = 500$ $\mu$m in water, with $D= 25.2$ mm and $V= 2.5$ m s$^{-1}$). (b) Rescaled data plotted against $\lambda \Delta \phi$ compared with the predicted scaling laws (thick dashed line) and the improved model (thin dashed lines, see Methods). (c) Pore pressure peak measured at different depth and rescaled by the predicted  scaling laws: $P_{f}^{max}\sim (\eta_{f} V D \alpha \Delta \phi/\kappa) \left( 6   A \alpha \eta_{f} D \Delta\phi /\rho_s \kappa V \right)^{-1/5}$ and $L=a\sim\sqrt{D V t_m}$ ($\Delta\phi=0.020$, same legend except for circles, for which: $V = 2.5$ m s$^{-1}$ and squares, for which: $V = 5.0$ m s$^{-1}$). } 
\label{fig:scaling}
\end{center}
\end{figure}
\begin{figure}
\begin{center}
\includegraphics[scale=1.2]{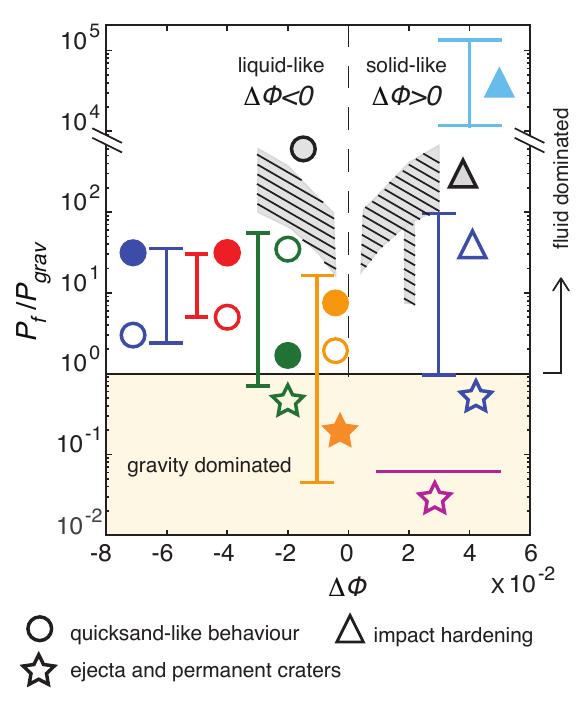}
\caption{An unified picture of impact responses in particulate media from dry granular materials to shear-thickening suspensions. Phase diagram ($P_f/P_{grav}$ vs $\Delta \phi$) showing diverse impact behaviours  (symbols) for a wide range of grain size ($d=10-500$ $\mu$m), impact velocity ($V=0.2-300$ m s$^{-1}$), fluid viscosity ($\eta_f=10^{-5}-10^{-2}$ Pa s) and fluid compressibility ($1/\chi=50-2.2\, 10^9$ Pa) with data from our work (dense suspensions, hatched area) and previous studies for glass beads in air \cite{Walsh_PRL2003} (orange), impacts at ultra-low air pressure mimicking planetary craters \cite{Yamamoto_2006} (purple), impacts on fine powders at atmospheric and low air pressure (green \cite{Royer_Nature2005}, red \cite{Caballero_PRL2007}, dark blue \cite{Royer_EPL2011}), impact on suspensions of cornstarch and water \cite{Waitukaitis_2012} (light blue).  Filled and open symbols correspond to incompressible and compressible case. The vertical bars give the range of $P_f/P_{grav}$ covered in the corresponding study and the horizontal bars give the uncertainty on $\Delta\phi$  (see Table in the Supplemental Material).}
%For $P_f/P_{grav}>1$, the pore--pressure dominates the impact dynamics and the response is predicted to be fluid-like ($\Delta \phi<1$) or solid-like ($\Delta \phi>1$) depending on the initial packing fraction. For $P_f/P_{grav}<1$, gravity dominates over the pore-pressure and the response is those of a dry granular material (ejecta and permanent craters). 

\end{center}
\label{fig:Model1}
\end{figure}

This dynamics is compared with experiments at various initial packing fraction $\phi_0>\phi_c$ for a given projectile and impact velocity in  Fig. S2 (see Supplemental Material).   As $\phi_0$ approaches $\phi_c$,  the indentation depth increases and the maximal deceleration decreases, while the stopping time increases. A collapse of all trajectories  is obtained when the variables are rescaled according to the model prediction (see Fig. S2 in Supplemental Material). Quantitative agreement  is achieved by taking $A \alpha\simeq 30$. Such a value is consistent with experimental measurements of $\alpha$ and $A$ for glass beads \cite{Pailha_JFM2009,Seguin_2013}, for which $\alpha \approx 4$ and $A\approx 5-10$. 
Furthermore, the Darcy-Reynolds model captures the impact dynamics for a wide range of physical parameters: using different suspension mixtures and different impact parameters, the data for the maximal indentation depth $\delta_{\rm max}$, stopping time $t_{\rm end}$ and maximal deceleration $\gamma_{max}$ collapse on the prediction when plotted as function of $\lambda \Delta \phi$ [Fig. 3(a,b)].  Finally, we systematically measure the pore--pressure profile below the impact point for different impactor parameters and  suspensions [Fig. 3(c)]. An universal exponential decay is obtained when the pore--pressure is scaled by the maximal pore--pressure given by the Darcy-Reynolds scaling (\ref{eq:DarcyReynoldsScaling}) and the depth $z$ is scaled by $L=a\sim\sqrt{D V t_m}$, thereby conclusively supporting the model.

These results can be easily extended to suspensions for which the interstitial fluid is not a liquid but a gas, like dry powders at different operating air pressure. As long as the diffusion timescale for gas expulsion $\tau_d\sim \eta \chi L^2/\kappa$, where $\chi$ is the gas compressibility, is small compared to the impact timescale $\tau_i\sim D/V$, the fluid can be assumed incompressible and the Darcy-Reynolds scaling (\ref{eq:DarcyReynoldsScaling}) of the pore--pressure still holds (see Supplemental Material).  Otherwise, the fluid has no time to escape from the pores during impact. The pore--pressure is then given by a gas state equation $P_f\sim -(1/\chi) \Delta \phi/(\phi_0(1-\phi_0))$, where $\chi\propto1/P_0$  and $P_0$ is the gas pressure.  In Fig. 4, we compare the predicted pore--pressure $P_f$ with the typical confining pressure due to gravity $P_{grav}\sim \phi_0 \Delta \rho g D$ for our study and previous impact studies covering a wide range of particulate media and impact conditions, for both incompressible and compressible interstitial fluids. When $P_f\gg P_{grav}$, the pore--pressure built-up during impact dominates the dynamics, yielding quicksand-like or solid-like response depending on the sign of $\Delta \phi$, in agreement with observations \cite{Royer_Nature2005,Caballero_PRL2007,Royer_EPL2011}. By contrast when $P_f\ll P_{grav}$ the interstitial fluid contributes negligibly to forces between grains and so, one recovers the classical dry granular case phenomenology \cite{Walsh_PRL2003,Yamamoto_2006}.    

Our study thus provides an unifying picture to explain the main regimes observed during the impact of a solid object onto a mixture of grains and fluid, at least when surface tension effects can be neglected (saturated suspension). It also provides a mechanism to explain the impact--activated solidification observed in more complex shear--thickening media like cornstarch \cite{Waitukaitis_2012}. In these systems, the critical packing fraction $\phi_c$ is expected to be a decreasing function of the impact velocity  \cite{Fernandez_2013,Seto_2013,Wyart_2014}, and not a material constant like in the present study. Thus, at high enough impact velocities, the suspension should become dilatant and solidify due to the pore--pressure feedback mechanism. Since the cornstarch particles are about 10 $\mu$m and the suspension permeability scale as $\kappa\propto d^2$, the Darcy-Reynolds scaling  (\ref{eq:DarcyReynoldsScaling}) predicts a 100-fold increases of the  pore--pressure compared to that of the Newtonian suspension with glass beads ($d=170$ $\mu$m). This is about $1-10$ MPa for typical impact conditions, which is in good agreement with impact stress estimations in cornstarch \cite{Waitukaitis_2012} [Fig. 4]. We confirm this viewpoint by performing qualitative measurements of pore--pressure in cornstarch suspensions. When an object is impulsively-moved in a concentrated suspension of cornstarch in water, pore--pressure takes indeed a huge drop (see Fig. S3 in Supplemental Material). Therefore, it is remarkable that the conjunction of two distinguished mechanisms in granular media, the Darcy law and the Reynolds dilatancy, could explain this long-standing puzzle: why we can run on cornstarch.

\vspace{5mm}
%\noindent \textbf{Supplementary information} is available in the online version of the paper.\\

\noindent\textbf{Acknowledgment} The authors thank Mathieu Leger and Sady No\"{e}l for the design of the experimental set--up. We also thank Emil Dohlen and students from Polytech'Marseille for preliminary results. This work was supported by the French National Research Agency (ANR) through the program No. ANR-11-JS09-005-01 and by the Labex MEC (ANR-10-LABX-0092) and the A*MIDEX project (ANR-11-IDEX-0001-02) funded by the ``Investissements d'Avenir'' French Government program.\\

%\noindent\textbf{Authors contribution} J.J.S.J., N.V. and Y.F conceived the study, performed experiments, built the model and wrote the paper. \\

\noindent\textbf{Author Information} The authors declare no competing financial interests. Correspondence and requests for materials should be addressed to Y.F. (yoel.forterre@univ-amu.fr).  

%%The authors thank Mathieu Leger and Sady No\"{e}l for the design of the experimental set--up. 
%This work was supported by the French National Research Agency (ANR) through the program No. ANR-11-JS09-005-01 and by the Labex MEC (ANR-10-LABX-0092) and the A*MIDEX project (ANR-11-IDEX-0001-02) funded by the ``Investissements d'Avenir'' French Government program. 

\section*{SUPPLEMENTAL MATERIAL}

\noindent\textbf{Material characterization.} The suspensions are prepared by mixing mono-dispersed glass beads of density $\rho_{p} = 2.50$ g  cm$^{-3}$ with a Newtonian liquid of density $\rho_f<\rho_p$ and viscosity $\eta_f$. Most of the experiments are conducted with a suspension of beads of diameter $d = 170$ $\mu$m (with a standard deviation of 20 $\mu$m) in water ($\rho_f=1.00$ g  cm$^{-3}$, $\eta_f=10^{-3}$ Pa s). In addition, suspension of larger glass beads of diameter $d = 500$ $\mu$m (with a standard deviation $60$ $\mu$m) immersed in water or in a mixture of UCON lubricant fluid (Dow, 75-H-90,000) and water ($\rho_f = 1.01$ g cm$^{-3}$, $\eta_f = 10.5 \times 10^{-3}$ Pa.s) are used to explore the influence of bead diameter and fluid viscosity.  Care was taken to use two batches of grains with similar internal friction coefficient (tested by comparing the angle of repose), in order to get similar value of the parameter $A$. The value of the critical volume fraction $\phi_c$ obtained by the pressure measurements  (change of sign of the peak pore-pressure after impact) in the case of $d=500$ $\mu$m is $\phi_c = 0.5940$ and is thus different from the other suspension (for which $\phi_c = 0.585$), probably due to differences in grain size distribution.

An important parameter of the model is the Darcy permeability $\kappa$ of the granular suspension, which relates the flow rate across the porous medium to the gradient of fluid pressure \cite{Jackson_Book2000}. The permeability of each suspension were obtained by imposing a downward gravity-driven flow through the grains and by measuring the corresponding flow rate, giving $\kappa = 3.0 \times 10^{-11}$ m$^2$ for the 170 $\mu$m diameter beads and $\kappa = 1.8 \times 10^{-10}$ m$^2$ for the 500 $\mu$m diameter beads. The variation of the permeability with the initial volume fraction of the pile is less than 15 $\%$ in the range studied and neglected in the model. \\

\noindent\textbf{Suspension preparation.}
The particles are mixed with the working liquid and placed in a large container (side length: 18 cm, depth: 27 cm). At the beginning of each experiment, the granular bed is fluidized by injecting the liquid through a porous plate placed at the bottom of the container. When the tank is almost filled with the liquid, the liquid supply is stopped and the particles are allowed to sediment. The tank wall is then gently tapped to compact the bed. The tapping on the side of the container is performed by a pendulum, released from a constant height, to ensure repeatability of the compacting process. Before each impact, the granular bed height $h_0$ is measured at each side of the container with an accuracy of 0.5 mm for a range of 9 to 10 cm depending on the number of taps. The initial volume fraction is deduced from these measurements with an accuracy of $\pm 0.002$. The excess liquid is later drained-off so that a thin liquid layer is kept on  the surface of the pile ($\sim 1$ mm).  The exact amount of water left does not affect the experimental results presented in this study. Finally, a solid sphere made of glass (density $\rho_s=2.52$ g cm$^{-3}$, diameter $D = 25.2$ or $16.5$ mm) is let to freely--fall on the granular bed, impacting at a speed $V$. A high-speed video camera records the dynamics (up to 150,000 frames per seconds).\\

%negligible (less than for the volume fraction considered in the study. haFor an assembly of spheres, a classical empirical law is the Carman-Kozeni formula: $\kappa=B [(1-\phi)^3/(150\phi^2)]d^2$, where $B=150-180$ is a constant.  the permeability is given by the    is given by related the flow rate and the gradient of pressure varies with the square of the particle size : $\kappa\propto d^2$.  
%
%by measuring the flow rate of a liquid of known velocity through a tube filled with the grains. and permeability $\kappa = 1.8 \times 10^{-10}$ m$^2$ the grain placed in a tube in a tube filled with the grains and measuring the flow rate. the pile driven by gravity through measuring the flow rate of a liquid of known viscosity through a tube filled with the grain ,  is an important property of the granular medium. The value $\kappa = 3.0 \times 10^{-11}$ m$^2$ is obtained by measuring the flow rate of a liquid of known velocity through a tube filled with the grains. and permeability $\kappa = 1.8 \times 10^{-10}$ m$^2$. k 

%%
\noindent\textbf{Pore--pressure measurements.}
To limit the perturbation induced by a sensor in the region of the suspension affected by impact, a low footprint pore--pressure sensor was designed. A thin, $15$ cm long, L--shaped stainless steel tube (internal diameter: $2$ mm, thickness: $0.5$ mm), ended by a fine metallic grid preventing the grains from entering into the tube is attached to a water-proof chamber. The tube and the chamber are filled with the working liquid. A hydrophone (Br\"{u}el \& Kj\ae r $8103$) with a flat frequency response in the range 1 Hz-15 kHz is used to measure the high frequency pressure signal within the chamber. An additional low frequency differential pressure sensor (Honeywell) is used to determine the slow relaxation dynamics. 
Because of the fine grid and of the complex geometry of the pore--pressure sensor, a careful calibration of this measuring device had to be conducted through a specific experiment described in Extended Data Fig. 1. The calibration chamber is divided into three compartments: compartment (I) contains the suspension to be impacted, compartment (II) contains a similar suspension but with a different height and finally compartment (III) is a water-filled chamber containing the pore--pressure sensor and a bare hydrophone separated from compartment (I) by a porous plate allowing transmission of the pressure signal. The purpose of compartment (II) is to ensure that the thin tubes are fastened just as the tubes in the experiment of the main text. After impact under conditions similar to the main experiment, the transfer function of the pore--pressure sensor was determined by measuring the signals from the bare hydrophone (BH) and the embedded hydrophone  (EH) of the pore--pressure sensor at different grain sizes and fluid viscosities.\\

\noindent\textbf{Refined impact model.}
The model presented in the main text can be slightly extended to offer a more accurate description of the dynamics when the indentation is not small and for larger values of $\Delta \phi$. The proposed modifications lie in the following two arguments:
\begin{enumerate} 
\item In the main text  model the radius of contact was approximated by $a \approx (\delta D)^{1/2}$, valid for small $\delta$. When the indentation $\delta$ is not small the radius of contact is better described by the following form: $a= D f(\delta/D)$ with $f(x) =  \sqrt{x - x^2}$ if $x < 1/2$ and $ f(x) = 1/2$ otherwise.
\item  In the main text  model the parameter $A$, relating the normal stress $\sigma_{zz}$ acting of the ball and the confining granular pressure $P_p$, was assumed to be constant. However, this parameter changes with the internal friction angle of the pile $\beta$, which itself depends on the initial volume fraction $\Delta \phi$ \cite{Andreotti_Forterre_Pouliquen_2013}.  To account for this dependence of $A$ with $\Delta \phi$, we model $A$ by the classical expression used in soil mechanics to describe the bearing capacity of a frictional soil (the so-called load-bearing capacity factor) \cite{Schofield_1968}, that is
\begin{equation}
A = \left( \frac{1 + \sin \beta}{1 - \sin \beta} \right) \exp ( \pi \tan \beta)
\end{equation}
and we relate the internal friction angle $\beta$ with $\Delta \phi$ using the dilatancy angle  \cite{Andreotti_Forterre_Pouliquen_2013}, so that  
$\tan \beta = \tan \beta_0 + \tan\Psi$, where $\beta_0$ is the internal friction angle at the critical state $\phi=\phi_c$ and $\tan\Psi=\alpha \Delta \phi$ as before.
\end{enumerate} 
%\begin{enumerate}
%\item In the main text the radius of contact was approximated by $a \approx (\delta D)^{1/2}$, valid for small $\delta$. When the indentation $\delta$ is not small the radius of contact is better described by the following form: $a= D f(\delta/D)$ with $f(x) =  \sqrt{x - x^2}$ if $x < 1/2$ and $ f(x) = 1/2$ otherwise.
%\item In the main text the parameter $A$, relating the normal stress $\sigma_{zz}$ acting of the ball and the confining granular pressure $P_p$, was assumed constant. However, this parameter is function of internal friction angle of the pile $\beta$, which itself depends on the initial volume fraction $\Delta \phi$ \cite{Andreotti_Forterre_Pouliquen_2013}.  To account for this dependence of $A$ with $\Delta \phi$, we model $A$ by the classical expression used in soil mechanics to describe the bearing capacity of a frictional soil (the so-called load-bearing capacity factor) \cite{Schofield_1968}, that is
%\begin{equation}
%A = \left( \frac{1 + \sin \beta}{1 - \sin \beta} \right) \exp ( \pi \tan \beta)
%\end{equation}
%and we relate the internal friction angle $\beta$ with $\Delta \phi$ using the dilatancy angle  \cite{Andreotti_Forterre_Pouliquen_2013}, so that  
%$\tan \beta = \tan \beta_0 + \tan\Psi$, where $\beta_0$ is the internal friction angle at the critical state $\phi=\phi_c$ and $\tan\Psi=\alpha \Delta \phi$ as before. 
%\end{enumerate}

With these two modifications, the equation for the dynamics of the sphere reads
\begin{equation}
\frac{d^2 \tilde{\delta}}{d \tilde{t}^2} = - \frac{A}{A_0} \left( \lambda \Delta \phi \right)^{3/5} \frac{d \tilde{\delta}}{d \tilde{t}} f \left[ \left( \lambda \Delta \phi \right)^{-2/5} \tilde{\delta} \right],
\end{equation}
where $A_0 = A|_{\Delta \phi =0}$ and $ \tilde{\delta} = \delta \left( \lambda \Delta \phi \right)^{2/5} / D$ and $\tilde{t} = V t \left( \lambda \Delta \phi \right)^{2/5} / D$, with
\begin{equation}
\lambda = \frac{6 \eta_f \alpha A_0}{\rho_b \kappa} \, \frac{D}{V}.
\end{equation} 
This model yields results that depend both on $\lambda$ and $\lambda \Delta \phi$. The predictions from this model with $\beta_0 = 25 ^o$ and $\alpha = 2.5$ are shown in Fig. 3 of the main text.\\

\noindent\textbf{Extension to compressible interstitial fluids.} In the model presented in the text, the interstitial fluid is assumed incompressible. When fluid compressibility is taken into account,  the linearized  mass conservation equation for the fluid phase becomes: $\rho_f\chi (1-\phi) \partial P_f/\partial t -\rho_f \partial\phi/\partial t +\rho_f (1-\phi) \nabla \cdot \textbf{V}_f  = 0$, where $\chi=(1/\rho_f)(\partial \rho_f/\partial P_f)$ is the compressibility of the fluid. Using as before the mass conservation of the solid phase and taking the divergence of the Darcy law (\ref{eq:Darcy}) gives:
\begin{equation}
\label{eq:Darcy-reynolds-compressible}
	\frac{1}{\phi} \frac{\partial \phi}{\partial t} = -\frac{\kappa}{\eta_{f}} \nabla^2 P_{f} + \chi (1-\phi) \frac{\partial P_f}{\partial t}=-\alpha  \dot{\gamma} \left({\phi-\phi_c}\right).  
	\end{equation}
When the diffusion time scale  $\tau_d\sim \eta_f \chi D^2/\kappa$ is short compared to the impact timescale $\tau_i\sim D/V$, the compressible term $\chi (1-\phi) \partial P_f/\partial t$ is small compared to the Darcy term $({\kappa}/{\eta_{f}}) \nabla^2 P_{f}$ and one recovers the incompressible Darcy-Reynolds equation (1-4) for the pore--pressure.  By contrast, when $\tau_d\gg\tau_i$, the compressible term is dominant and the pore-pressure is given by: $(1/{\phi}) {\partial \phi}/{\partial t} = \chi (1-\phi) \partial P_f/\partial t$, that is $P_f\sim -(1/\chi) \Delta \phi/(\phi(1-\phi))$. In Fig 4, the pore--pressure is computed as $P_{f}=  (\eta_{f}/\kappa)  \vert\Delta \phi\vert V D$ when the compressible number $C=\tau_d/\tau_i<1$ and as   $P_f= (1/\chi) \vert\Delta \phi\vert/(\phi_0(1-\phi_0))$ when $C>1$.

%\clearpage 

%\noindent \textbf{Supplementary Video 1.} Fluid-like response for $\phi_0<\phi_c$. Movie of a solid sphere ($D=25.2$ mm, speed 2.35 m s$^{-1}$) impacting a non-buoyant suspension in the loose case (glass beads of diameter $d=170$ $\mu$m in water, initial volume fraction $\phi_0=0.56$). The real duration of the sequence is 300 ms; the movie is slow-down 100 times. Movie corresponding to Fig. 1b upper panel.   \\
%
%\noindent \textbf{Supplementary Video 2.} Solid-like response $\phi_0>\phi_c$. Same condition as in Movie S1 except that the initial volume fraction is $\phi_0=0.604$ (dense case).  The real duration of the sequence is 60 ms; the movie is slow-down 100 times. Movie corresponding to Fig. 1b lower panel. \\
%
%\noindent \textbf{Supplementary Video 3.} Role of grain diameter and fluid viscosity in the impact response for $\phi_0>\phi_c$. In the 3 sequences,  the speed of the impacting sphere ($D=25.2$ mm) is $V=2.55$ m s$^{-1}$ and $\Delta \phi=\phi_0-\phi_c=0.047$.  Left: suspension of glass beads of diameter $d=170$ $\mu$m in water ($\eta_f=10^{-3}$ Pa s).  Center:  glass beads of diameter $d=1300$ $\mu$m in water. Right: glass beads of diameter $d=1300$ $\mu$m in silicon oil ($\eta_f=10^{-1}$ Pa s).  The real duration of the sequences is 130 ms; the movie is slow-down 100 times. \\
%
%\noindent \textbf{Supplementary Video 4.} Impact in a dry granular material (glass beads of diameter $d=170$ $\mu$m in air, $\phi_0=0.62$, $D=16.5$ mm, $V=3.0$ m s$^{-1}$).   The real duration of the sequences is 253 ms; the movie is slow-down 100 times. 

\begin{figure*}
\begin{center}
\includegraphics[scale=1.2]{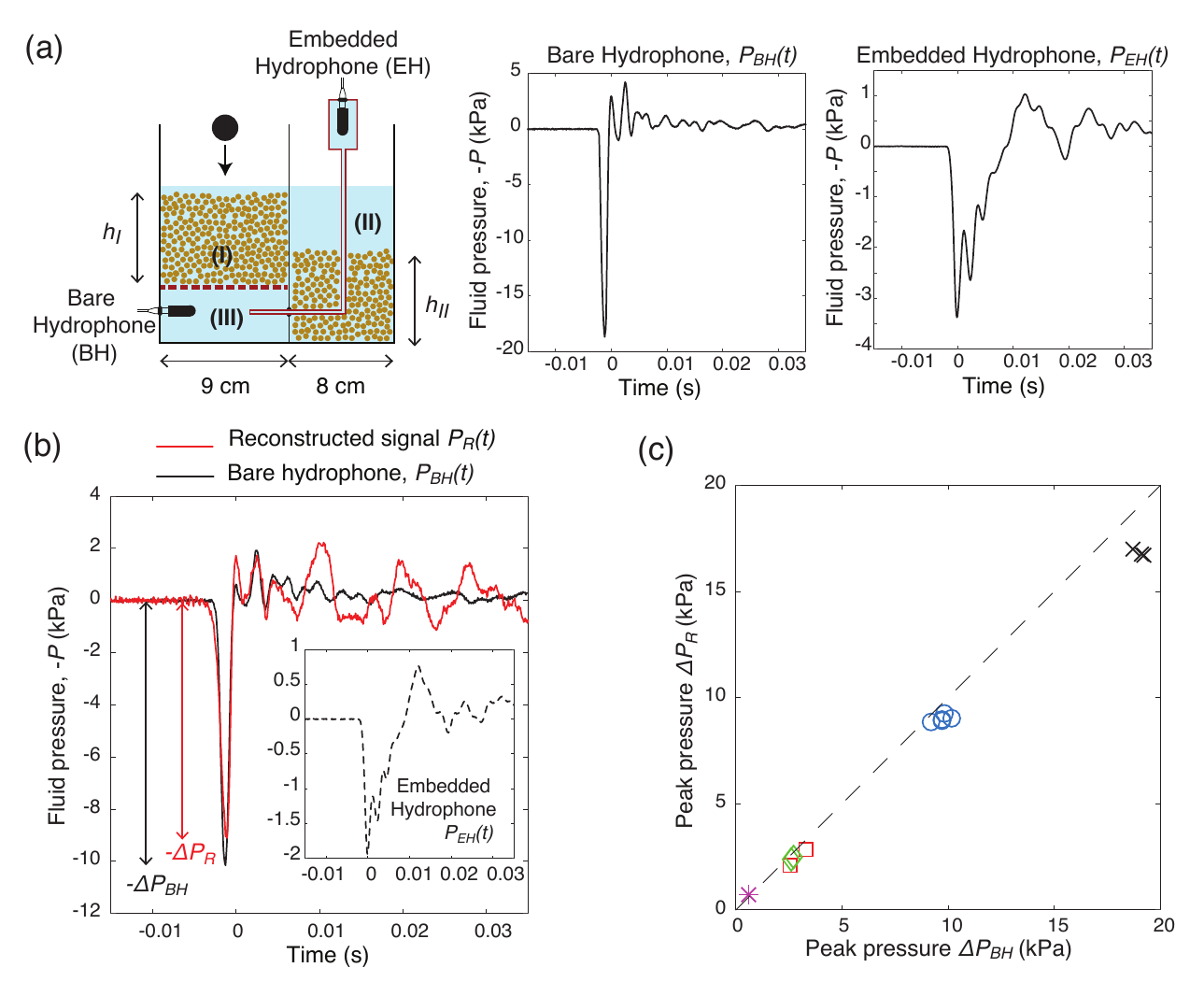}
\end{center}
\vspace{0.5cm} 
\noindent FIG. S1: Calibration procedure for the pore--pressure measurement. (a) Calibration chamber ($h_{I}\sim 9$ cm, $h_{II}\sim 6.5$ cm) and signal $P_{BH}(t)$ (resp. $P_{EH}$) recorded by the bare hydrophone (resp. by the embedded hydrophone) during a typical impact condition with $\phi_0=0.61$, $V=5.9$ m s$^{-1}$, $D=25.2$ mm,  $d=500$ $\mu$m, $\eta_f=10.5$ Pa s. The embedded hydrophone (grid and tube) induces a strong distorsion in both in the amplitude and temporal characteristics of the input signal. The inverse transfer function of the sensor in the Fourier space, $\tilde{G}_{inv}$, is obtained from the signals $P_{BH}(t)$ and $P_{EH}(t)$ using the following formula: $\tilde{G}_{inv}=1/{\rm FFT}(G\times f)$, where $G={\rm iFFT}[{\rm FFT}(P_{EH})/{\rm FFT}(P_{BH})]$ and $f(t)=(1-\tanh((t-t_0)/dt))/2$ is a filtering function used to enforce causality. The inverse transfer function is then averaged over 3 trials performed at   the same impact condition and used for all calibrations made in the study.  (b) Red: Input signal reconstructed from the signal of the embedded hydrophone (inset) using the inverse transfer function $\tilde{G}_{inv}$ determined previously, for a different impact velocity ($V=4.1$ m s$^{-1}$). The reconstructed signal $P_{R}$ is obtained from the signal of the embedded hydrophone $P_{EH}$ using the formula: $P_{R}={\rm iFFT}[\tilde{G}_{inv}\times {\rm FFT}(P_{EH})]$. Black: input signal $P_{BH}$ measured directly by the bare hydrophone for the same impact. The transfer function introduces spurious oscillations on long times in the reconstructed signal but correctly reproduces the peak pressure just after impact.   (c) Comparison between the peak pressure measured with the bare hydrophone $\Delta P_{BH}$ and the peak pressure of the reconstructed signal $\Delta P_R$ for different impact conditions (black crosses: $V=5.9$ m s$^{-1}$; blue circles: $V=4.1$ m s$^{-1}$; red squares:  $V=1.2 $ m s$^{-1}$; green diamonds: $V=1.2$ m s$^{-1}$, $h_{II}\sim 3.5$ cm;  purple star: $V=1.2$ m s$^{-1}$, impact in a looser pile). In all cases, the reconstructed peak pressure is within 20 $\%$ of the real peak pressure. 
\label{figsupp1}
\end{figure*}

%\cleardoublepage

%
\begin{figure*}[h!]
\begin{center}
\includegraphics[scale=1.2]{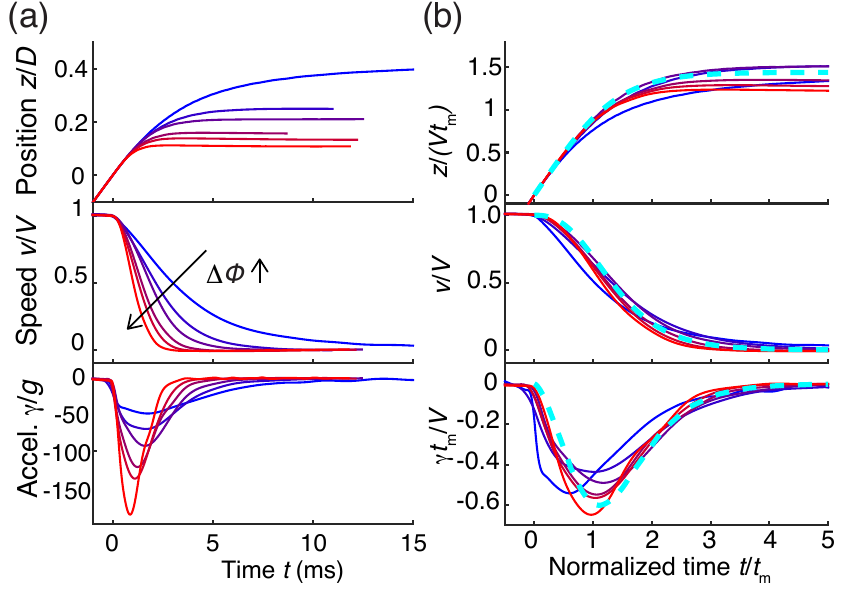}
\end{center}
\vspace{0.5cm} 
\noindent FIG. S2: Penetration dynamics  in the dense regime. (a) Experimental position, speed and acceleration for $\Delta \phi = 0.001$, $0.004$, $0.007$, $0.01$, $0.013$, $0.019$ (glass beads $d = 170$ $\mu$m in water, $D = 25.2$ mm, $V = 2.5$ m s$^{-1}$). (b) Same data rescaled using the characteristic time $t_m$ and length $V t_m$ together with the model prediction (dashed line).  
\label{fig:Model1}
\end{figure*}

\begin{figure*}[h!]
\begin{center}
\includegraphics[scale=1.2]{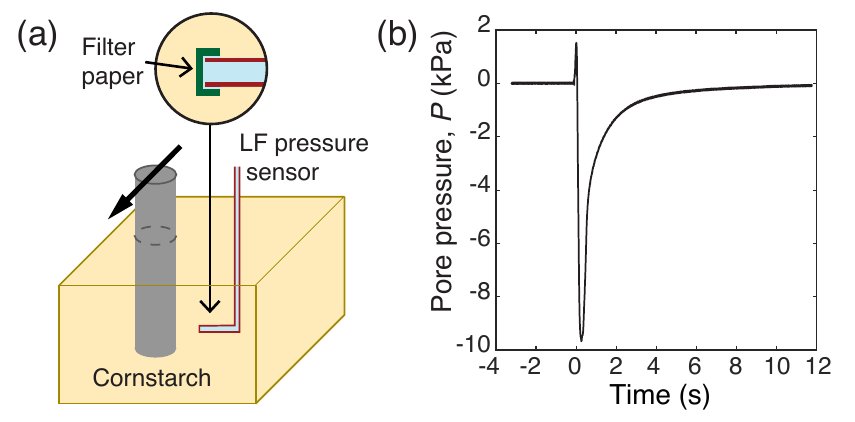}
\end{center}
\vspace{0.5cm} 

\noindent FIG. S3: Transient hardening is associated with negative pore--pressure in dense suspensions of cornstarch in water. (a) A cylinder is impulsively moved by hand  and the pore--pressure is recorded simultaneously. The pressure sensor is connected to a tube filled with water and ended by a fine semi-permeable paper that prevents cornstarch particles to enter the tube.  (b) Pore--pressure versus time (cornstarch 55 $\%$ wt in distilled water, the suspension has been vigorously stirred before the experiment to avoid sedimentation effects). The cylinder is put in motion at $t=0$.   
\label{fig:scaling}
\end{figure*}

\begin{figure*}[h!]
\begin{center}
\includegraphics[scale=0.9]{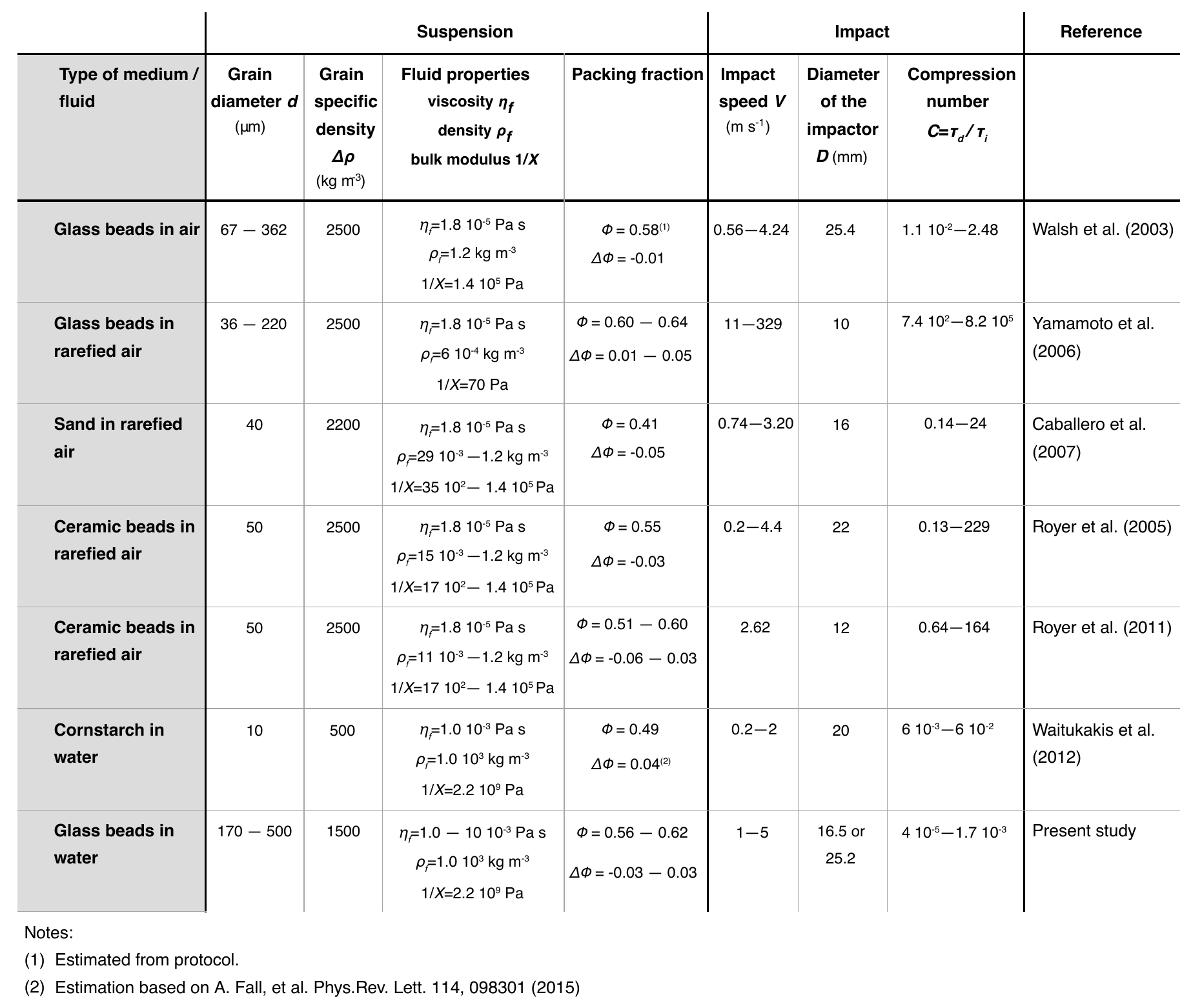}
\end{center}
\vspace{0.5cm} 

\noindent TAB. S1: Impact parameters and granular/fluid properties used to plot the phase diagram in Fig. 4. For each material, the permeability $\kappa$ is computed using Carman-Kozeni formula \cite{Jackson_Book2000}:  $\kappa=(1-\phi)^3 d^2/(180\phi^2)$. The air compressibility is given by $\chi=1/(\gamma P_0)$, where $P_0$ is the air pressure and $\gamma\simeq 1.4$.   
\label{fig:table}
\end{figure*}

\end{document}